\documentclass[twocolumn,pr,showpacs,notitlepage,nobibnotes,nofootinbib]{revtex4-1}
\usepackage{amssymb}
\usepackage{amsmath}
\usepackage{graphicx}
\usepackage{bm}
\usepackage{color}
\usepackage[T2A]{fontenc}
\usepackage[cp1251]{inputenc}
\usepackage[english]{babel}

\newcommand{\eps}{\varepsilon}

\newcommand{\braket}[3]{\left\langle #1 \left| #2 \right| #3 \right\rangle}

\newcommand{\ket}[1]{\left| #1 \right\rangle}

\newcommand{\A}{\mathcal A}
\newcommand{\B}{\mathcal B}
\newcommand{\D}{\mathcal D}
\newcommand{\e}{\mathrm e}
\renewcommand{\phi}{\varphi}

\begin{document}

\title{Effects of electron-hole asymmetry on electronic structure of helical edge states in HgTe/HgCdTe quantum wells}
\author{M. V. Durnev\footnote{durnev@mail.ioffe.ru}}
\affiliation{Ioffe Institute, 194021 St. Petersburg, Russia}

\begin{abstract}

We study the effects of electron-hole asymmetry on the electronic structure of helical edge states in HgTe/HgCdTe quantum wells. In the framework of the four-band \textit{\textbf{kp}}-model, which takes into account the absence of a spatial inversion centre, we obtain analytical expressions for the energy spectrum and wave functions of edge states, as well as the effective $g$-factor tensor and matrix elements of electro-dipole optical transitions between the spin branches of the edge electrons.  We show that when two conditions are simultaneously satisfied — electron-hole asymmetry and the absence of an inversion centre — the spectrum of edge electrons deviates from the linear one, in that case we obtain corrections to the linear spectrum.

\end{abstract}

\maketitle
\section{Introduction}

The study of helical edge states emerging at the edge of two-dimensional topological insulators is an important area of physics of two-dimensional crystalline systems with nontrivial topological properties~\cite{Bernevig15122006, Konig:2007it, PhysRevLett.107.136603}. The key areas of research include the study of local and nonlocal electron transport through edge channels~\cite{Roth17072009, PhysRevB.84.121302, Ma:2015aa, PhysRevLett.123.056801}, backscattering mechanisms~\cite{PhysRevLett.106.236402, PhysRevB.86.035112, PhysRevLett.111.086401, PhysRevB.90.115309, Entin2015, Kurilovich2017, doi:10.1002/pssr.201700422} and photogalvanic effect~\cite{PhysRevB.95.201103, doi:10.1002/andp.201800418}. Among various systems, where one-dimensional helical channels are experimentally observed~\cite{Konig:2007it, PhysRevLett.107.136603, Fei:2017qq}, HgTe/HgCdTe quantum wells attract the most attention. In such wells, the transition between the trivial and topological phases occurs upon variation of the quantum well width.

Electronic states in HgTe/HgCdTe quantum wells of close-to-critical width are usually obtained in the framework of the four-band \textit{\textbf{kp}}-model, which includes closely lying electron and hole subbands. The isotropic modification of this model is called Bernevig-Hughes-Zhang (BHZ) model~\cite{Bernevig15122006} and is widely used for calculations of the electronic structure of the bulk and edge states~\cite{RevModPhys.83.1057, Konig:2008fk, PhysRevLett.101.246807, PhysRevB.82.113307, PhysRevB.91.035310, enaldiev2015, PhysRevB.92.155424, Entin_2017}.
However, as shown by the atomistic calculations~\cite{PhysRevB.77.125319, PhysRevB.91.081302}, the absence of a center of spatial inversion in the zinc blende lattice and the low symmetry of the HgTe/HgCdTe quantum well heterointerfaces lead to strong mixing of the electron and hole subbands, which modifies the \textit{\textbf{kp}}-model~\cite{Konig:2008fk, Winkler20122096, PhysRevB.91.081302} and leads to a substantial rearrangement of bulk and edge electronic states~\cite{PhysRevB.91.081302, PhysRevB.93.075434, PhysRevB.93.155304}. The absence of a center of spatial inversion in the HgTe/HgCdTe quantum well is responsible, for example, for the emergence of optical transitions between helical states with opposite spin in the framework of the strong electro-dipole mechanism~\cite{0953-8984-31-3-035301}.

Both in the isotropic BHZ model and in the noncentrosymmetric \textit{\textbf {kp}}-model, there are diagonal terms proportional to the square of the wave vector and related to the presence of remote energy bands. These contributions are different for electron and hole subbands, which leads to violation of electron-hole symmetry -- the Hamiltonian does not coincide with itself after replacing an electron with a hole and simultaneously changing the sign of the energy. Although the terms quadratic in wave vector do not directly affect the group velocity and localization width of the edge states (these quantities are controlled by the large inter-subband mixing, which is linear in wave vector), electron-hole asymmetry encoded in diagonal terms, leads to significant modification of the energy spectrum and wave functions of edge states. Electron-hole asymmetry may also result in new effects. For instance, it is responsible for the appearance of circular photogalvanic effect in helical edge channels~\cite{PhysRevB.92.155424, PhysRevB.95.201103, entin_magarill_2016, 0953-8984-31-3-035301}. The influence of electron-hole asymmetry on the spectrum and structure of edge states has been studied earlier in the framework of the BHZ model (see, for example, works~\cite{PhysRevB.92.155424, Entin_2017}). However, such studies were not carried out for realistic quantum wells that do not have a center of spatial inversion.

In this paper we study the effects of electron-hole asymmetry on helical edge states in HgTe/HgCdTe quantum wells. In particular, we analyze the dispersion of edge electrons and derive expressions for the wave functions of edge states in the framework of noncentrosymmetric \textit{\textbf{kp}}- model, we calculate the components of the $g$-factor tensor of edge electrons and the matrix elements of optical transitions between the spin branches of edge channel. The paper is organised as follows: in Sec.~\ref{sec:zero_ky} a general model is presented and expressions for the energy and wave functions of the edge states are obtained for a zero wave vector along the edge; in Sec.~\ref{zeeman} the $g$-factor tensor of edge electrons is calculated; in Sec.~\ref{nonzero_ky} the spectrum and wave functions of edge states are found in a wide range of wave vectors; in Sec.~\ref{optical} the matrix elements of optical transitions between the spin branches of edge states are calculated; and finally, in Sec.~\ref{boundary} we discuss the effect of boundary conditions on the obtained results.

\section{Edge states at zero wave vector} \label{sec:zero_ky}

In this section we formulate the model and obtain wave functions of helical states at zero wave vector of the motion along the edge. Let us consider a HgTe/CdHgTe quantum well grown along the [001] crystallographic direction with a width, which is close to the critical width, when the transition to a topological phase occurs. Such a structure possesses $D_{2d}$ point symmetry lacking the center of a spatial inversion. The states in the vicinity of the Fermi level are formed from the close-in-energy electron-like and hole-like subbands $\ket{E1,\pm 1/2}$ and $\ket{H1, \pm 3/2}$, respectively~\cite{Bernevig15122006}. In the $\ket{E1, +1/2}$, $\ket{H1, +3/2}$, $\ket{E1, -1/2}$, $\ket{H1, -3/2}$ basis the quantum well states are described by the following Hamiltonian~\cite{PhysRevB.93.075434}:
\begin{widetext}
\begin{equation}
\label{eq:H_bulk}
\mathcal H_0(k_x,k_y) =
\left( 
\begin{array}{cccc}
\delta_0 - (\B+\D)k^2 & {\rm i} \A k_+ & 0 & {\rm i} \gamma \e^{-2\mathrm{i}\theta} \\
-{\rm i} \A k_- & - \delta_0 + (\B-\D)k^2 & {\rm i} \gamma \e^{-2\mathrm{i}\theta} & 0\\
0 & -{\rm i} \gamma \e^{2\mathrm{i}\theta} & \delta_0 - (\B+\D)k^2 & - {\rm i} \A k_- \\
-{\rm i} \gamma \e^{2\mathrm{i}\theta} & 0 & {\rm i} \A k_+ & - \delta_0 + (\B-\D)k^2
\end{array}
\right) \: .
\end{equation}
\end{widetext}
Here $\bm k = (k_x, k_y)$ is the electron in-plane wave vector, $k = |\bm k|$, $k_\pm = k_x \pm \mathrm{i} k_y$, $\A$, $\B$, $\D$, $\delta_0$ and $\gamma$ are real band structure parameters. At $\B < 0$ ($\B > 0$) and $\delta_0 < 0$ ($\delta_0 > 0$) the quantum well is in the topological insulator phase, and its' edges support helical edge states. The $\gamma$ parameter takes into account the absence of a spatial inversion center in the structure, and results mainly from the mixing of $\ket{E1, \pm1/2}$ and $\ket{H1, \mp3/2}$ subbands at the quantum well interfaces~\cite{PhysRevB.91.081302}. The value of this parameter is not yet determined experimentally, and theoretical values lie in the range $2 \div 5$~meV for a HgTe/Hg$_{0.3}$Cd$_{0.7}$Te quantum well~\cite{Konig:2008fk, PhysRevB.77.125319, Winkler20122096, PhysRevB.91.081302}.

\begin{figure}[htpb]
\includegraphics[width=0.35\textwidth]{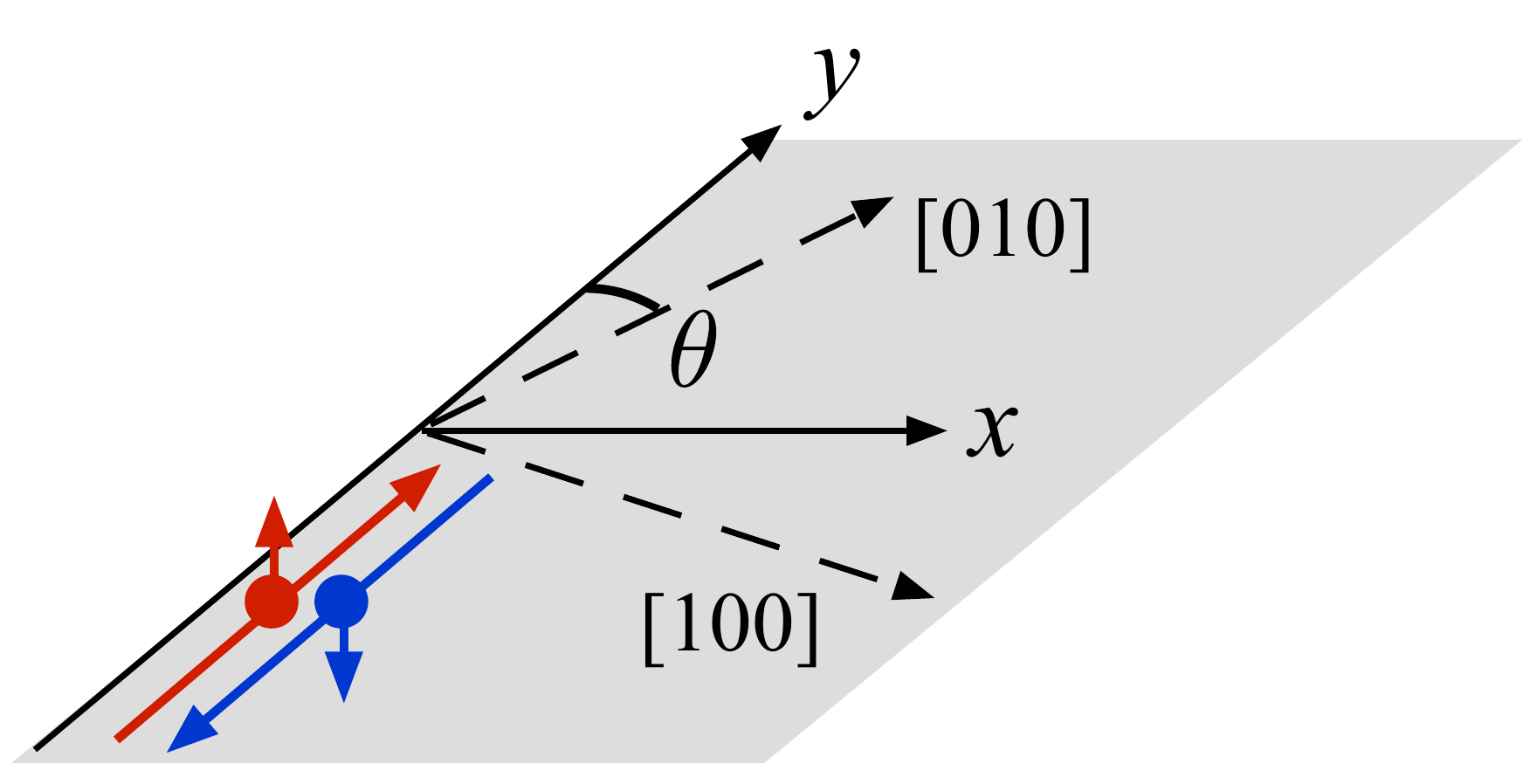}
\caption{\label{fig:edge} The edge of topological insulator based on HgTe/CdHgTe quantum well with a sketch of helical edge states.
}
\end{figure}

In what follows, we use a coordinate system with the $z$-axis that coincides with the growth axis of the quantum well, the edge of the sample is parallel to the $y$ axis, and the sample occupies the half-space $x>0 $, see Fig.~\ref{fig:edge}. The $y$ axis makes the angle $\theta$ with the [010] crystallographic axis, which allows one to consider structures with different crystallographic orientations of the edge. In particular, at $\theta = 0$ the edge of the sample is parallel to the [010] axis, while at $\theta = \pi/4$ the edge is parallel to the [110] axis. While obtaining analytical expressions and analysing the results, we assume that $\B <0$, $\delta_0 <0$, $|\D | < |\B|$, $\A> 0$ and $\gamma> 0$. For numerical estimates, we use the set of parameters $\A = 3.6$~eV$\cdot$\AA, $\B = -68$~eV$\cdot$\AA$^2$, $\D = -51$~eV$\cdot$\AA$^2$~\cite{Konig:2008fk}, $\gamma = 5$~meV~\cite{PhysRevB.91.081302}, and $\delta_0 = -10$~meV corresponding to a HgTe/Hg$_{0.3}$Cd$_{0.7}$Te quantum well with a 8 nm width.

Diagonal terms in the Hamiltonian~\eqref{eq:H_bulk}, proportional to $k^2$, arise due to the mixing of the four considered subbands with remote subbands that are not included in the Hamiltonian. This mixing is described by the parameters $\B$ and $\D$. As seen, in the case of $\D \neq 0$ the diagonal terms enter asymmetrically for the electron and hole subbands, and therefore break the symmetry of the Hamiltonian with respect to the replacement of an electron by a hole. Hence, the ratio $\D/\B$ can be regarded as the electron-hole asymmetry parameter. As will be shown below, the spectrum and wave functions of the edge states will be largely determined by this parameter. For the mentioned parameters, $\D/\B \approx 0.75$.

The edge states are derived from the solution of the Schr\"odinger equation
\begin{equation}
\label{Schr}
\mathcal H_0 \left(-\mathrm{i} \frac{\partial}{\partial x}, k_y \right) \psi_{k_y s} = \eps_{k_y s} \psi_{k_y s}
\end{equation} 
with the boundary conditions $\psi_{k_y s}(x = 0) = 0$ и $\psi_{k_y s}(x \to +\infty) = 0$. For each value of $k_y$ there exist a pair of such states with different pseudospin $s = \pm 1/2$. We use here boundary conditions of the simplest form, other types of boundary conditions are discussed in Sec.~\ref{boundary}. 

First, let us consider $k_y = 0$. We will now show that in this case there exist an analytical solution of Eq.~\eqref{Schr} with an energy
\begin{equation}
\label{eps0}
\eps_{0s} = -\delta_0 \frac{\D}{\B}\:.
\end{equation}
For this purpose we substitute energy~\eqref{eps0} in Eq.~\eqref{Schr} and write the sought wave functions in the form
\begin{equation}
\label{wfs0}
\psi_{0 +1/2} = \frac{\e^{\mathrm{i} k_y y}}{\sqrt{2 L}} \left[
\begin{array}{c}
a(x) \\
-\alpha a(x) \\
-\mathrm{i} b(x) \e^{2 \mathrm{i} \theta}  \\
-\mathrm{i} \alpha b(x) \e^{2 \mathrm{i} \theta}
\end{array}
\right]\:,
\end{equation}
\begin{equation*}
\psi_{0 -1/2} = \frac{\e^{\mathrm{i} k_y y}}{\sqrt{2 L}} \left[
\begin{array}{c}
- \mathrm{i} b(x) \e^{-2 \mathrm{i} \theta} \\
\mathrm{i} \alpha b(x) \e^{-2 \mathrm{i} \theta} \\
a(x) \\
\alpha a(x)
\end{array}
\right]\:,
\end{equation*}
where
\begin{equation}
\alpha = \sqrt{\frac{\B + \D}{\B - \D}}\:.
\end{equation}
After such an ansatz the equations for  $a(x)$ and $b(x)$ are analogous to the equations at $\D = 0$. Using the solution at $\D = 0$~\cite{PhysRevB.93.075434} we obtain
\begin{eqnarray}
\label{ab}
a(x) &=& \mathcal N \left[ \e^{-x/l_1} \cos \frac{\phi}{2} - \e^{-x/l_2} \cos \left( k_0 x  - \frac{\phi}{2} \right) \right]\:, \nonumber \\
b(x) &=& \mathcal N \left[ \e^{-x/l_1} \sin \frac{\phi}{2} + \e^{-x/l_2} \sin \left( k_0 x  - \frac{\phi}{2} \right) \right]\:,
\end{eqnarray}
where 
\begin{equation}
\mathcal N = \frac{2}{\sqrt{l_2 (1 + \alpha^2)}}\:,
\end{equation}
\begin{equation}
\label{lengths}
l_1 = - \frac{\varkappa \B}{\A}\:,~l_2 =  -\frac{\A}{\varkappa \delta_0} \:,~k_0 = \frac{\gamma}{\A}\:,\: \tan \phi = -\frac{\gamma}{\varkappa \delta_0}\:,
\end{equation}
and
\begin{equation}
\label{kappa}
\varkappa = \frac{\sqrt{\B^2 - \D^2}}{|\B|}\:.
\end{equation}

To derive~\eqref{ab} we used the relation $l_1 \ll l_2$, which is valid for realistic quantum wells. Indeed, it follows from~\eqref{lengths}, that $l_1 \approx 10$~\AA, $l_2 \approx 540$~\AA~for above mentioned parameters.

In what follows we assume that $l_1 \ll l_2$ relation is fulfilled. Equations~\eqref{eps0}-\eqref{kappa} 
generalise results of Ref.~\cite{entin_magarill_2016} ($\D \neq 0$, $\gamma = 0$) and Ref.~\cite{PhysRevB.93.075434} ($\D = 0$, $\gamma \neq 0$). It is seen from the derived relations, that despite the small values of  
$\B$ and $\D$ (which manifests itself in small $l_1$), all the results depend significantly on $\D/\B$. 

The key consequence of electron-hole asymmetry is the shift of Dirac point ($\eps_{0 s}$) from the middle of the gap ($\eps = 0$). As seen from~\eqref{eps0}, the Dirac point shift is proportional to $\D/\B$: at $\D < 0$ the Dirac point shifts towards the conduction band, whereas at $\D > 0$ -- towards the valence band. The shift of the Dirac point leads to a redistribution of the relative contributions of the subbands $\ket{E1,\pm 1/2}$ and $\ket{H1,\pm 3/2}$ to the wave functions, increase of the decay length  $l_2$, and change of the edge electrons velocity. In the limiting cases $\D = \pm \B$, the Dirac point falls on the boundary of the energy gap, and the edge states at $k_y = 0$ hybridise with the bulk ones. As follows from~\eqref{lengths}, in this case $l_2 \to \infty$.

By projecting the part of the Hamiltonian~\eqref{eq:H_bulk} with $k_y$ and $k_y^2$ terms onto the wave functions~\eqref{wfs0} we obtain the energy of the edge electrons up to linear in $k_y$ terms:
\begin{equation}
\label{Hedge}
\eps_{k_y s} = -\delta_0 \frac{\D}{\B} + 2s \hbar v_0 k_y\:,
\end{equation}
where the edge velocity
\begin{equation}
\label{velocity}
v_0 = \frac{\A}{\hbar} \frac{|\delta_0| \varkappa^2}{\sqrt{\varkappa^2 \delta_0^2 + \gamma^2}}\:.
\end{equation}

\section{Zeeman effect for edge states} \label{zeeman}

In the framework of \textit{\textbf{kp}}-model the interaction of electron with magnetic field $\bm B$ 
is described by the sum of the orbital contribution, given by the Hamiltonian~\eqref{eq:H_bulk} with the Peirels substitution $\mathcal H_0 [\bm k - (e/c \hbar) \bm A]$, where $\bm A$ is the vector potential of magnetic field, and the Zeeman contribution~\cite{PhysRevB.93.075434}
\begin{equation}
\label{Hz}
\mathcal H_Z = \frac{\mu_B}{2} \left(
\begin{array}{cccc}
g_e^\perp B_z & 0 & g_e^\parallel B_- & 0 \\
0 & g_h^\perp B_z & 0 & g_h^\parallel \e^{-4\mathrm{i}\theta} B_+ \\
g_e^\parallel B_+ & 0 & -g_e^\perp B_z & 0 \\
0 & g_{h}^\parallel \e^{4 \mathrm{i} \theta} B_- & 0 & -g_h^\perp B_z
\end{array}
\right)\:,
\end{equation}
where $g_e^\parallel$, $g_e^\perp$, $g_h^\parallel$ and $g_h^\perp$  are the $g$-factors of $\ket{E1}$ and $\ket{H1}$ subbands, that contain contributions from the mixing with remote electron and hole subbands, $B_\pm = B_x \pm \mathrm{i} B_y$, and $\mu_B$ is the Bohr magneton.

The interaction of edge electrons with magnetic field is described by the following effective Hamiltonian in the basis $(\psi_{0+1/2},\psi_{0-1/2})$:
\begin{equation}
\mathcal H_{\rm edge}^{(\bm B)} = \frac{\mu_B}{2} \sum \limits_{\alpha, \beta = x,y,z} g_{\alpha \beta} \sigma_\alpha B_\beta\:,
\end{equation}
where $g_{\alpha \beta}$ are the components of the $g$-factor tensor of edge electrons, and $\sigma_x, \sigma_y, \sigma_z$ are the Pauli matrices.
By projecting the Hamiltonian~\eqref{Hz} onto the wave functions~\eqref{wfs0}, we obtain the components of the $g$-factor tensor for $\bm B$ lying in the quantum well plane
\begin{eqnarray}
g_{xx} &=& g_1 \cos^2 2 \theta + g_2 \sin^2 2\theta\:, \\
g_{yy} &=& g_1 \sin^2 2\theta + g_2 \cos^2 2 \theta\:, \nonumber \\
g_{xy} &=& g_{yx} = \frac12 (g_1 - g_2) \sin 4 \theta\:, \nonumber
\end{eqnarray}
where
\begin{eqnarray}
\label{g12}
g_1 &=& \frac12 \left( \frac{\B - \D}{\B} g_e^\parallel - \frac{\B + \D}{\B} g_h^\parallel \right)\:, \\
g_2 &=& \frac12 \left( \frac{\B - \D}{\B} g_{e}^\parallel + \frac{\B + \D}{\B} g_{h}^\parallel \right) \frac{|\delta_0| \varkappa}{\sqrt{\varkappa^2 \delta_0^2 + \gamma^2}}\:. \nonumber
\end{eqnarray}

Equation~\eqref{g12} generalises results of~\cite{PhysRevB.93.075434} for $\D \neq 0$. Since the heavy-hole $g$-factor $g_h^\parallel$ is close to zero in quantum wells made of materials with a zinc-blende lattice~\cite{Mar99}, the main contribution to $g_1$ and $g_2$ is given by the first terms in~\eqref{g12}. In real structures, the $(\B - \D)/\B$ factor may strongly deviate from unity, for example, for the parameters listed in Sec.~\ref{sec:zero_ky}, $(\B - \D)/\B \approx 1/4$, hence the $g$-factors for the in-plane magnetic field are decreased by about four times as compared with $\D = 0$. Estimates give $g_{1} \approx 2.5$, $g_2 \approx 2$, consistent with the results of numerical modeling of the edge electrons spectrum in~\cite{0953-8984-31-3-035301}.

Magnetic field directed normal to the quantum well mixes the $\ket{E1}$ and $\ket{H1}$ subbands leading to to large orbital contribution to the $g_{\alpha z}$ components. The diagonal component $g_{zz}$, similarly to the case of $\D = 0$, depends on the vector potential gauge and can be set to be zero~\cite{PhysRevB.93.075434}, while the off-diagonal components are gauge-independent and have the following form:
\begin{equation}
g_{xz} = -g_3 \sin 2\theta,~~~g_{yz} = g_3 \cos 2 \theta\:,
\end{equation}
where
\begin{eqnarray}
g_3 = \frac{2 m_0 \A^2}{\hbar^2} \frac{\gamma |\delta_0| \varkappa^2}{(\delta_0^2 \varkappa^2 + \gamma^2)^{3/2}}\:.
\end{eqnarray}
Estimations give $g_3 \approx 130$, and thus, $g_3 \gg g_1,~g_2$.

Magnetic field mixes edge states and opens a gap in its spectrum. This energy gap at $\theta = 0$ is
\begin{equation}
\label{Eg_B}
\eps_g = \mu_B \sqrt{g_1^2 B_x^2 + (g_2 B_y + g_3 B_z)^2}\:.
\end{equation}
As follows from~\eqref{Eg_B}, the gap opens for any direction of the field, except when the field lies in the $yz$-plane and is directed at an angle $-\arctan g_2/g_3$ with respect to $y$-axis. We note, that for any edge orientation (any $\theta$) there exists a direction in space, when applied magnetic field does not open a gap.

Giant anisotropy of the $g$-factor is confirmed in the magneto-transport experiments. It was shown that perpendicular magnetic field suppresses edge conductivity due to the opening of the Zeeman gap, whereas the influence of the in-plane magnetic field on the conductivity is much weaker~\cite{PhysRevLett.123.056801} .

\section{Edge states at nonzero wave vector} \label{nonzero_ky}

In order to study electron transport and optical transitions involving edge electrons, it is necessary to know the wave functions and the spectrum of edge states for a nonzero wave vector of motion along the edge. In this section, we find the spectrum and wave functions of edge states for $k_y\neq 0$. In the framework of the isotropic model corresponding to $\gamma = 0$ in the Hamiltonian~\eqref{eq:H_bulk}, and also in the framework of the model without an inversion center, but possessing electron-hole symmetry ($\gamma \neq 0$, $\D = 0$), analytical expressions are derived for the spectrum and wave functions in the entire range of wave vectors. In a more general case, approximate results are obtained that are valid for a small value of $\D/\B$ or $\gamma/|\delta_0|$.

\subsection{Isotropic model} \label{sec:iso}

First, let us consider the case $\gamma = 0$, which corresponds to the isotropic model. Detailed study of the edge states structure within isotropic approximation is presented in~\cite{Entin_2017}. In this case the Hamiltonian~\eqref{eq:H_bulk} is composed of two independent 2$\times$2 blocks. 
At $\gamma = 0$ the Schr\"odinger equation~\eqref{Schr} has an analytical solution, valid for a wide range of $k_y$~\cite{PhysRevLett.101.246807, Entin_2017}
\begin{equation}
\label{eps_edge}
\eps_{k_y s}^{(0)} = -q \delta_0 + 2 s \varkappa \A k_y\:,
\end{equation}
\begin{equation}
\label{wfs_ky_iso}
\psi_{k_y +1/2}^{(0)} = \frac{\e^{\mathrm{i} k_y y}}{\sqrt{2 L}} \left[
\begin{array}{c}
a(x, k_y) \\
-\alpha a(x, k_y) \\
0  \\
0
\end{array}
\right]\:,
\end{equation}
\begin{equation*}
\psi_{k_y -1/2}^{(0)} = \frac{\e^{\mathrm{i} k_y y}}{\sqrt{2 L}} \left[
\begin{array}{c}
0 \\
0 \\
a(x, -k_y) \\
\alpha a(x, -k_y)
\end{array}
\right]\:.
\end{equation*}
Here $q = \D/\B$, function $a$ has the form [see Eq.~\eqref{ab} at $\gamma = 0$]
\begin{equation}
\label{aky}
a(x, k_y) = \frac{2}{\sqrt{l_2(k_y) (1 + \alpha^2)}} \left[ \e^{-x/l_1} - \e^{-x/l_2 (k_y)} \right]\:,
\end{equation}
with the edge states localization width $l_2$ that depends on the wave vector
\begin{equation}
\label{l2ky}
l_2(k_y)^{-1} = l_2^{-1} - q k_y - k_y^2 l_1\:,
\end{equation}
and $l_1$ and $l_2$ are given by Eqs.~\eqref{lengths}.
The derived expression for $a(x, k_y)$ is valid, as before, at $l_1 \ll l_2 (k_y)$. The $k_y^2 l_1$ term is small in the range of the studied  $k_y$, and thus, is neglected in the following. Important consequence of the electron-hole asymmetry ($q \neq 0$) is the dependence of the edge localization length on $k_y$. As shown below, this dependence leads to nonzero matrix element of the electric dipole operator between the edge states.

Let us consider the case $q > 0$ in more detail.  As follows from Eq.~\eqref{l2ky}, the length $l_2(k_y)$ diverges at $k_y = k^* = 1/(q l_2)$ for the edge state with $s = +1/2$ and at $k_y = -k^*$ for the $s = -1/2$ state. One may show that at $k_y = \pm k^*$ dispersion curves of the edge states~\eqref{eps_edge} touch the lower boundary of the conduction band $\eps_{\rm c} = \sqrt{\delta_0^2 + \A^2 k_y^2}$. At these points the edge states ``merge'' with the bulk (2D) states. However, note that the elastic scattering from the edge states to the conduction band comes to play significantly earlier, already at $\eps_{k_y s} = - \delta_0$, i. e., $k_y =  1/l_2(1+q)$. While moving along the edge branches towards the valence band the $l_2(k_y)$ length decreases. At $\eps_{k_y s} = \delta_0$, when the energy of the edge states coincides with the maximum energy of the valence band one has $l_2(k_y) = l_2(1-q)$. At subsequent increase of the wave vector $l_2(k_y)$ may become comparable with $l_1$, and Eqs.~\eqref{aky}, \eqref{l2ky} are not valid in this case. This situation is analyzed in detail in Ref.~\cite{Entin_2017}.

\subsection{Model without inversion center}

Let us now consider the model of realistic quantum wells lacking the center of spatial inversion. This model is described by the Hamiltonian~\eqref{eq:H_bulk} with $\gamma \neq 0$. We will derive analytical results in two limits, in the limit $\gamma/|\delta_0| \ll 1$, when anti-diagonal terms of the Hamiltonian~\eqref{eq:H_bulk} can be considered as a small perturbation, and in the limit $|\D/\B| \ll 1$, i.e., in the limit of weak electron-hole asymmetry.
 
 \subsubsection{The limit $\gamma/|\delta_0| \ll 1$} \label{weak_gamma}

In this section we consider anti-diagonal terms of the Hamiltonian~\eqref{eq:H_bulk} as a small perturbation. Non-perturbed wave functions $\psi_{k_y \pm 1/2}^{(0)}$ of the edge states are found in Sec.~\ref{sec:iso} for arbitrary $k_y$, see Eq.~\eqref{wfs_ky_iso}. Note that in the first order of perturbation theory over $\gamma$ the perturbation does not couple or shift in energy the edge states $\psi_{k_y \pm 1/2}^{(0)}$. It means that the corrections to the edge states wave functions are due to the admixture of bulk states of conduction or valence band to the edge states.

Hence, in the first order of perturbation theory over $\gamma$ the energy of edge states is not changed and coincides with Eq.~\eqref{eps_edge}. 
We will seek the wave functions of the edge states in the following form
\begin{equation}
\label{wfs_ky_aniso}
\psi_{k_y +1/2} = \frac{\e^{\mathrm{i} k_y y}}{\sqrt{2 L}} \left[
\begin{array}{c}
a_1 (x,k_y) \\
-\alpha a_2 (x,k_y) \\
-\mathrm{i} b_1 (x, k_y) \e^{2 \mathrm{i} \theta} \\
-\mathrm{i} \alpha b_2 (x, k_y) \e^{2 \mathrm{i} \theta}
\end{array}
\right]\:,
\end{equation}
\begin{equation*}
\psi_{k_y -1/2} = \frac{\e^{\mathrm{i} k_y y}}{\sqrt{2 L}} \left[
\begin{array}{c}
- \mathrm{i} b_1 (x, -k_y) \e^{-2 \mathrm{i} \theta} \\
\mathrm{i} \alpha b_2 (x, -k_y) \e^{-2 \mathrm{i} \theta} \\
a_1 (x, -k_y) \\
\alpha a_2 (x, -k_y)
\end{array}
\right]\:,
\end{equation*}
where $b_1$ and $b_2$ are unknown functions $\propto \gamma$. 
In the first order of perturbation theory $a_1$ and $a_2$ are equal and coincide with the non-perturbed wave function~\eqref{aky}.

 The function $\psi_{k_y -1/2}$ is related to $\psi_{k_y +1/2}$ by the time inversion.  Eq.~\eqref{Schr} with non-perturbed energy~\eqref{eps_edge} leads to the following set of equations on $b_1$ and $b_2$:
\begin{eqnarray}
l_1 b_1 '' + b_2 ' + \left[ l_2(k_y)^{-1} + k_y \right] b_1 + k_y b_2 &=&  - k_0 a\:, \\
l_1 b_2 '' + b_1 ' + \left[ l_2(k_y)^{-1} - k_y \right] b_2 - k_y b_1 &=&  - k_0 a\:. \nonumber
\end{eqnarray}
The solution of these equations that satisfies the boundary conditions has the form
\begin{equation}
\label{b1b2}
b_1 = b(x) + B(x)\:,~~ b_2 = b(x) - B(x)\:,
\end{equation}
where
\begin{multline}
\label{bB}
b(x) = \frac{ \mathcal N(k_y) k_0 l_2(k_y) }{2} \times \\ 
\times \left\{ \e^{-x/l_1} + \left[ \frac{2x}{l_2(k_y)} - 1 \right] \e^{-x/l_2(k_y)} \right\}\:, \\ 
B(x) = - \mathcal N(k_y) k_0 k_y l_2(k_y) ~x \e^{-x/l_2(k_y)}\:,
\end{multline}
and
\[
\mathcal N(k_y) = \frac{2}{\sqrt{l_2(k_y) (1 + \alpha^2)}}\:.
\]
At $k_y = 0$ the functions $b_1$ and $b_2$ are equal and coincide with Eq.~\eqref{ab}, taken in the limit $\gamma/|\delta_0| \ll 1$.
The perturbation theory is valid when $b_{1,2}$ are small compared to $a$, i.e. at $k_0l_2(k_y) \ll 1$. It limits the range of $k_y$, over which the derived equations can be applied.

Using corrections to the wave functions, obtained in the first order of perturbation theory over $\gamma/|\delta_0|$, we will derive the second order corrections to the edge states energy:
\begin{equation}
\eps_{k_y s}^{(2)} = \braket{\psi_{k_y s}^{(0)}}{V_\gamma}{\psi_{k_y s}^{(1)}}\:,
\end{equation}
where $\psi_{k_y s}^{(0)}$ are the wave functions of the zero order~\eqref{wfs_ky_iso}, $\psi_{k_y s}^{(1)}$ is the linear-in-$\gamma$ part of the wave functions~\eqref{wfs_ky_aniso}, and $V_\gamma$ is the anti-diagonal part of the Hamiltonian~\eqref{eq:H_bulk}. Calculations show that 
\begin{equation}
\label{eps_second}
\eps_{k_y s}^{(2)} = -\frac{\gamma^2}{2 |\delta_0|} \frac{k_y l_2}{(1 - q k_y l_2)^2}\:.
\end{equation}
Equation~\eqref{eps_second} contains quadratic in $\gamma$ correction to the edge velocity, as well as, for $q \neq 0$, the terms of higher powers in $k_y$. Hence, when two conditions are simultaneously fulfilled, electron-hole asymmetry and the absence of spatial inversion center ($q \neq 0$ and $\gamma \neq 0$), the spectrum of edge electrons deviates from the linear one. As before, Eq.~\eqref{eps_second} is valid when $k_0l_2(k_y) \ll 1$, i.e. at $1 - q k_y l_2 \gg k_0 l_2$.

\begin{figure}[htpb]
\includegraphics[width=0.48\textwidth]{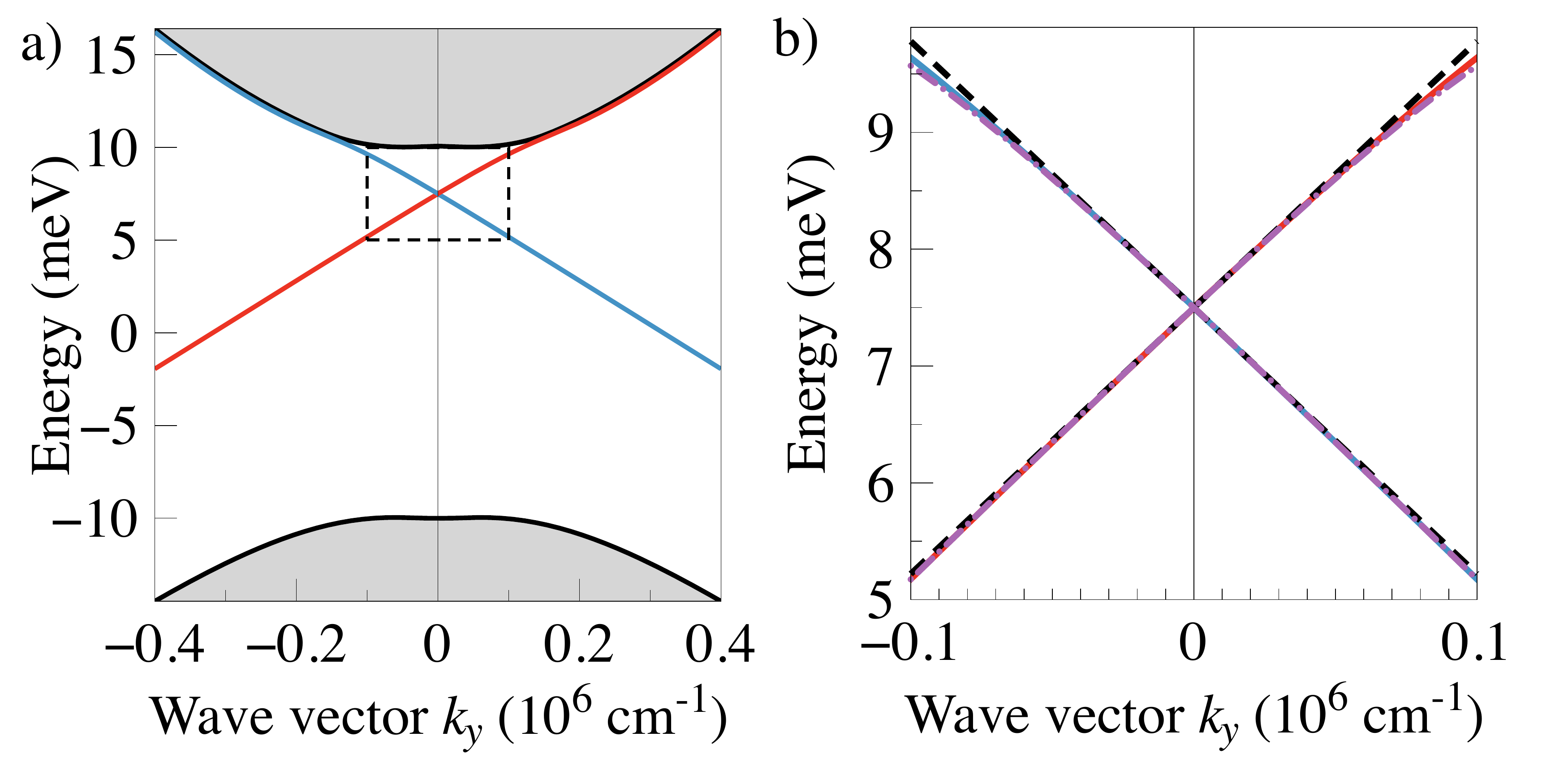}
\caption{\label{fig1} Energy spectrum of electronic states in HgTe/HgCdTe quantum well for band parameters shown in the text and $\gamma = 2$~meV, so that $\gamma/|\delta_0| = 0.2$. The edge dispersion branches are shown by blue and red. Panel b) shows increased part of the edge dispersion. Solid lines are numeric calculations, dashed lines are linear dependences with velocity calculated by~\eqref{velocity}, dashed-dotted lines are dependences $\eps_{k_y s}^{(0)} + \eps_{k_y s}^{(2)}$, which take into account deviation from the linear behaviour.
}
\end{figure}

Figure~\ref{fig1} shows the results of numerical calculation of edge and bulk states in the HgTe/CdHgTe well and comparison with the obtained analytical dependences. It can be seen that the spectrum of edge electrons deviates from the linear one with a velocity~\eqref{velocity}, and this deviation is well described by the dependence $\eps_{k_y s} = \eps_{k_y s}^{(0)} + \eps_{k_y s }^{(2)}$.
Corrections to the linear dispersion of edge electrons were also studied in~\cite{Entin_2017} in the framework of the isotropic model, but with boundary conditions of a more complex form. The nonlinearity of the spectrum leads, for example, to the generation of edge photocurrent in the edge channels due to indirect optical transitions~\cite{doi:10.1002/andp.201800418}.

\subsubsection{The limit of weak electron-hole asymmetry}\label{low_D}

For arbitrary $\gamma$, but $q = \D/\B \ll 1$ it is also possible to obtain analytical expressions for edge states. As before, we will seek the wave functions in the form~\eqref{wfs_ky_aniso}. Let us consider $\psi_{k_y +1/2}$ in more detail. The wave function $\psi_{k_y -1/2}$ is related to $\psi_{k_y +1/2}$ by Eq.~\eqref{wfs_ky_aniso}, and $\eps_{k_y -1/2} = \eps_{-k_y +1/2}$. The Schr\"odinger equation~\eqref{Schr} for $\psi_{k_y +1/2}$ leads to the following set of equations
\begin{eqnarray}
\label{sys}
l_1 a_1'' + a_2' + \left( l_2^{-1} + \frac{E}{\alpha \A} \right) a_1 - k_y a_2 - k_0 b_2 = 0\:, \nonumber \\
l_1 a_2''+ a_1' + \left( l_2^{-1} -  \frac{ \alpha E}{\A} \right) a_2 + k_y a_1 - k_0 b_1 = 0\:, \nonumber \\
l_1 b_1'' + b_2 ' + \left( l_2^{-1} + \frac{E}{\alpha \A} \right) b_1 + k_y b_2 + k_0 a_2 = 0\:,  \nonumber \\
l_1 b_2'' + b_1 ' + \left( l_2^{-1} -  \frac{\alpha E}{\A} \right) b_2 - k_y b_1 + k_0 a_1 = 0\:,
\end{eqnarray}
where we decomposed the edge state energy as $\eps_{k_y +1/2} = -q \delta_0 + E$.

The set of equations~\eqref{sys} has analytical solution at $q = 0$ ($\alpha = 1$) with energy
\begin{equation}
\label{eps_ky_gamma}
E^{(0)} = \frac{\A k_y |\delta_0|}{\sqrt{\delta_0^2 + \gamma^2}} \:.
\end{equation}
Thus, the spectrum of edge states obtained in Sec.~\ref{sec:zero_ky} using perturbation theory for small $k_y$ remains linear in the case $q = 0$ in the entire range of wave vectors. Details of the solution, as well as expressions for the functions $a_{1,2}$ and $b_{1,2}$ are given in the Appendix.

At $q \ll 1$ to the first order in $q$ we have $\alpha \approx 1 + q$ and $\alpha^{-1} \approx 1 - q$. Hence, Eqs.~\eqref{sys} have the same form as at $q = 0$ with $l_2$ that depends on energy
\begin{equation}
\label{l2_E}
l_2^{-1}(E) = l_2^{-1} - \frac{q E}{\A}\:.
\end{equation} 
In this equation up to terms linear in $q$ we can set $E = E^{(0)}$ and therefore obtain
\begin{equation}
\label{l2_gamma_ky}
l_2^{-1}(k_y) \approx l_2^{-1} - \frac{q k_y |\delta_0|}{\sqrt{\delta_0^2 + \gamma^2}}\:.
\end{equation} 
Equation~\eqref{l2_gamma_ky} describes the dependence of localization width of the $s = +1/2$ edge state on $k_y$ at $q \ll 1$.

To find corrections to the dispersion of the edge states~\eqref{eps_ky_gamma} at $q \ll 1$ we substitute the expression~\eqref{l2_gamma_ky} for $l_2 (k_y)$ in the right side of Eq.~\eqref{eps_ky_gamma} for energy. Up to the first order in $q$ we find
\begin{equation}
\label{quadratic}
E \approx E^{(0)} - q \A^2 k_y^2 \frac{\gamma^2 |\delta_0|}{(\delta_0^2 + \gamma^2)^2} \:.
\end{equation}
It follows from Eq.~\eqref{quadratic} that in the presence of electron-hole asymmetry the spectrum of edge states deviates from linear one,
and corrections quadratic in the wave vector appear. This result is consistent with that obtained in the limit $\gamma/|\delta_0| \ll 1$, see Eq.~\eqref{eps_second}.

\section{Matrix elements of optical transitions} \label{optical}

Excitation of the edge of topological insulator by electromagnetic wave causes optical transitions between the spin branches of edge states. Due to the absence of a spatial inversion center in HgTe/HgCdTe quantum wells, optical transitions occur due to electro-dipole mechanism~\cite{0953-8984-31-3-035301}. The matrix element of the electron-photon interaction is proportional to the matrix elements of the velocity operator $\bm v = \partial \mathcal H_0/ \hbar \partial \bm k$ between the states $\psi_{k_y s}$ and $\psi_ {k_y -s}$~\cite{0953-8984-31-3-035301}:
\begin{eqnarray}
v^{(x)}_{s-s} &=& \braket{\psi_{k_y s}}{v_x}{\psi_{k_y -s}} = u_1 \e^{-4 \mathrm{i} s \theta}\:,\nonumber \\
v^{(y)}_{s-s} &=& \braket{\psi_{k_y s}}{v_y}{\psi_{k_y -s}}   = 2\mathrm{i} s u_2 \e^{-4 \mathrm{i} s \theta}\:.
\end{eqnarray}

In this section we calculate $u_1$ and $u_2$ in the limit $\gamma/|\delta_0| \ll 1$. Wave functions $\psi_{k_y s}$ for that case are found in Sec.~\ref{weak_gamma}, see Eqs.~\eqref{wfs_ky_aniso}, \eqref{b1b2}, \eqref{bB}.
The calculation gives:
\begin{eqnarray}
\label{u1calcs}
u_1 &=& \frac{v_0 k_0 k_y l_2^2 \left[ l_2(k_y) - l_2(-k_y) \right]}{2 \sqrt{l_2(k_y) l_2(-k_y)}}  \:, \\
u_2 &=& \frac{v_0 k_0 l_2^2}{\sqrt{l_2(k_y) l_2(-k_y)}} \left[ 1 - \frac{l_2(k_y) + l_2(-k_y)}{2 l_2} \right] \nonumber  \:,
\end{eqnarray}
where $v_0 = \varkappa \A/\hbar$ is the edge velocity at $\gamma/|\delta_0| \ll 1$.
After simplifications we obtain
\begin{equation}
\label{u12}
u_1 = v_0 \frac{q k_0 k_y^2 l_2^3}{\sqrt{1 - q^2 k_y^2 l_2^2}}\:,~~ u_2 = -q u_1\:.
\end{equation}
Equations~\eqref{u1calcs}, \eqref{u12} are valid for arbitrary value of $q$ and wave vector $k_y$, for which the relation $1 - q k_y l_2 \gg k_0 l_2$ holds. Also in calculations we set $l_1 = 0$ both in the wave functions and the velocity operator. As follows from~\eqref{u1calcs}, the matrix elements $u_1$ and $u_2$ are nonzero due to the dependence of the edge states localization width on $k_y$, and therefore vanish if the system possesses electron-hole symmetry ($q = \D/ \B = 0$). It can be shown, however, that, taking into account small contributions $\propto l_1/l_2$, the matrix element $u_2 \neq 0$ even at $q = 0$.

\begin{figure}[htpb]
\includegraphics[width=0.48\textwidth]{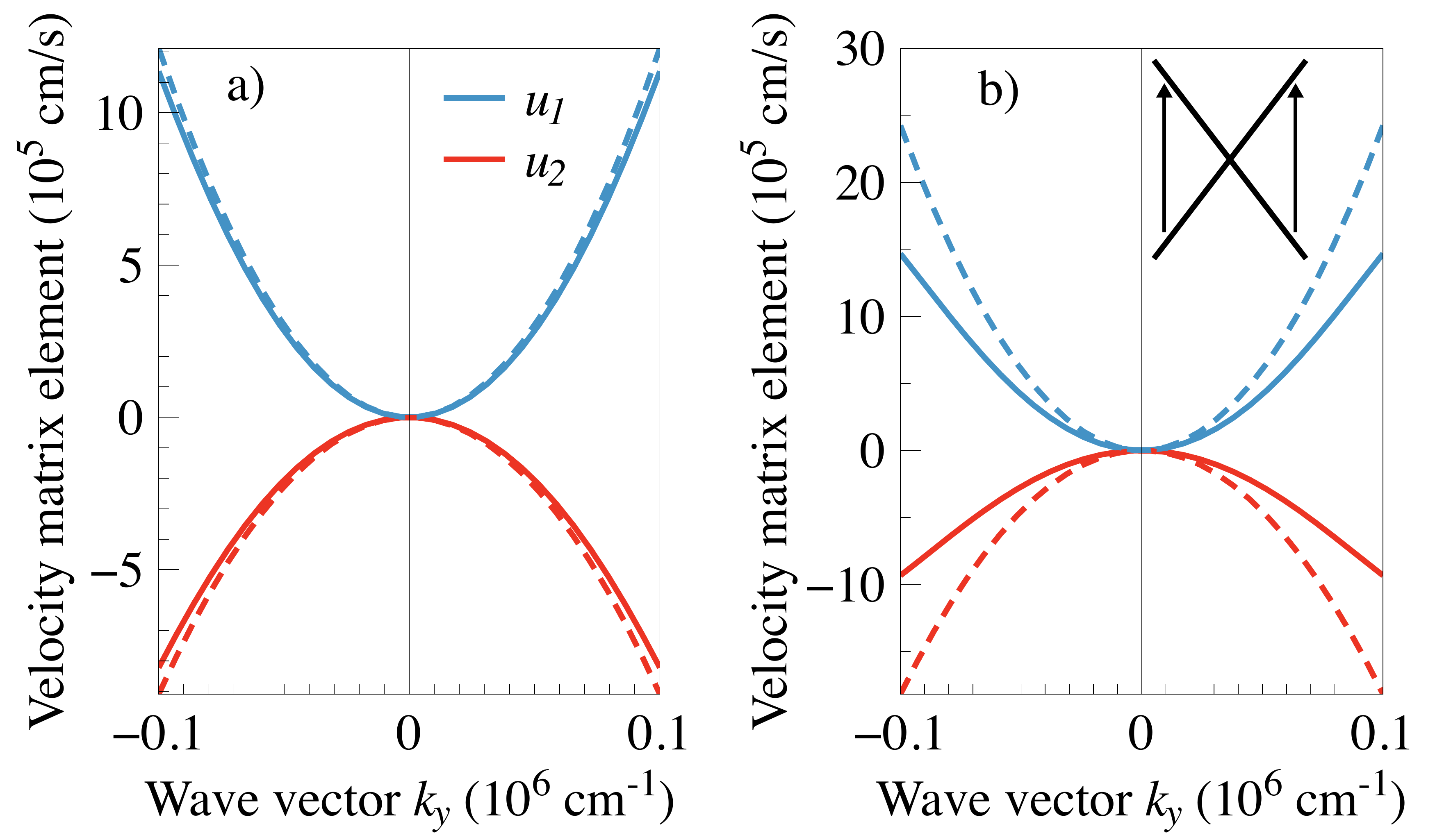}
\caption{\label{fig:fig2} Matrix elements of the velocity operator $u_1$ and $u_2$ for $\gamma/|\delta_0| = 0.1$ (а) and $\gamma/|\delta_0| = 0.2$ (b). Solid curves show numeric calculations, dashed curves show calculations by Eqs.~\eqref{u12}. The inset sketches the optical transitions in the system.
}
\end{figure}

Figure~\ref{fig:fig2} shows the results of calculations of the matrix elements $u_1$ and $u_2$ using two methods -- numerical diagonalization of the Hamiltonian~\eqref{eq:H_bulk} and analytical formulas~\eqref{u12}. The plots show that at $\gamma/|\delta_0| = 0.1$, the analytical calculations are in good agreement with the numerical ones, however at $\gamma/|\delta_0| = 0.2$, a significant discrepancy is already observed, so that Eqs.~\eqref{u12} overestimate the values of $u_1$ and $u_2$. One of the reasons is that with increasing $\gamma$  the edge velocity $v_0$, which enters expressions for $u_1$ and $u_2$, decreases.

Optical transitions between edge states can also be characterized by the matrix elements of the dipole moment operator, which at small $k_y$ have the form~\cite{0953-8984-31-3-035301}:
\begin{eqnarray}
d^{(x)}_{s-s} = -2s \mathrm{i} \e^{-4 \mathrm{i} s \theta} D_1 k_y\:,~~d^{(y)}_{s-s} = \e^{-4 \mathrm{i} s \theta} D_2 k_y\:,
\end{eqnarray}
where $D_{1,2} = eu_{1,2}/|k_y \omega_{s -s}|$ and $\hbar \omega_{s-s} = \eps_{k_y s} - \eps_{k_y -s}$. With account for Eq.~\eqref{u12}
\begin{eqnarray}
\label{D12}
D_1 &=& \frac{e}{2} q k_0 l_2^3 = \frac{e \D \B^2}{2 (\B^2 - \D^2)^{3/2}} \frac{\gamma \A^2}{\delta_0^3}\:, \\
D_2 &=& - q D_1 = -\frac{e \D^2 \B}{2 (\B^2 - \D^2)^{3/2}} \frac{\gamma \A^2}{\delta_0^3} \:. \nonumber
\end{eqnarray}
Estimations by Eq.~\eqref{D12} at $\gamma/|\delta_0| = 0.1$ give $D_1/e \approx 1.7\times10^{-12}$~cm$^2$ and $D_2/e \approx -1.2\times10^{-12}$~cm$^2$. This estimation agrees on the order of magnitude with the numeric estimations obtained in~\cite{0953-8984-31-3-035301} for $\gamma/|\delta_0| = 0.5$.

Upon absorption of the circularly polarized light by helical edge states, optical transitions occur asymmetrically in $k$ space leading to the generation of spin polarization and edge photocurrent~\cite{0953-8984-31-3-035301}. Relative difference in the rates of optical transitions from the states $\psi_{k_y -1/2}$ and $\psi_{-k_y +1/2}$, $g_{k_y -1/2}$ and $g_{-k_y +1/2}$, respectively, is proportional to the degree of circular polarization of the incident light $P_{\rm circ}$ and is equal to
\begin{equation}
\frac{g_{k_y -1/2} - g_{-k_y +1/2}}{g_{k_y -1/2} + g_{-k_y +1/2}} = KP_{\rm circ}\:,
\end{equation}
where
\begin{equation}
K = -\frac{2 D_1 D_2}{D_1^2 + D_2^2} = \frac{2 \D \B}{\B^2+\D^2}\:.
\end{equation}
We note that the asymmetry parameter $K$ for transitions between edge states is equal to the similar coefficient for transitions from edge to bulk states~\cite{PhysRevB.92.155424}. In both cases, $K = 0$ if the system possesses electron-hole symmetry ($\D = 0$). This result is a consequence of a more general statement about the absence of spin polarization and the photogalvanic effect upon the absorption of circularly polarized radiation by electron-hole symmetric system~\cite{0953-8984-31-3-035301}.

\section{The role of boundary conditions} \label{boundary}

The above results are obtained for the simplest “open” boundary condition $\psi(x= 0) = 0$. Solutions corresponding to a more general boundary condition $\psi'(x= 0) + h \psi(x=0) = 0$ differ from those considered in the work only by the pre-factors before the exponents $\e^{-x/l_1}$ and $\e^{-x/l_2}$ in the functions $a(x)$ and $b(x)$, see for example Eq.~\eqref{ab}. As a result, a change of $h$ parameter leads to a change of the wave functions exactly near the edge at the small $l_1$ scale, and therefore all the obtained results are independent of $h$, since they are determined by the wave functions behaviour on a much larger scale $l_2$.

The most general form of the boundary conditions at the boundary of the topological insulator with vacuum was obtained from general physical considerations in~\cite{enaldiev2015}. Boundary conditions of the general form take into account the admixture of remote subbands by the edge of the structure. In particular, it is possible that the boundary condition itself violates the electron-hole symmetry in the system. In this case, even at $\D = \B = 0$ in the quantum well Hamiltonian, the spectrum of edge states becomes asymmetric with respect to the center of the band gap and deviates from the linear~\cite{enaldiev2015, Entin_2017}.

Another example of a system with helical states is the boundary of two HgTe/HgCdTe quantum wells of different widths in the phase of trivial and topological insulators, respectively. In the simplest case, such a contact can be modelled by the spatial dependence $\delta_0(x)$ in the Hamiltonian~\eqref{eq:H_bulk}, assuming that the remaining band parameters are weakly dependent on the well width. In the case $\B = \D = 0$, such a model predicts the presence of edge states with a symmetric linear spectrum~\cite{PhysRevD.13.3398, volkov_pankratov_1985}. However, if $\B \neq 0$ and $\D \neq 0$, the situation becomes more complicated. The Dirac point position and the edge velocity depend on the ratio of $\delta_0$ values at the right and left sides of the contact. Let us consider $\delta_0(x>0) = -\delta_r$, $\delta_0(x<0) = \delta_l$, and $\delta_{l, r} >0$. Then, if the relation $\A/\delta_r,~\A/\delta_l \gg l_1$ holds, the electron-hole asymmetry does not lead to a noticeable change in the spectrum of edge states. In this case, the $\propto k^2$ diagonal terms in the Hamiltonian can be neglected without changing the physical results. The same situation is realized in the case of a smooth contact. However if $\A/\delta_r \gg l_1$ and $\A/\delta_l \lesssim l_1$, the electron-hole asymmetry begins to play a significant role in the spectrum of edge states. In the limit $\A/\delta_l \ll l_1$ the wave function does not penetrate into the region $x <0$, which corresponds to the open boundary condition.

\section{Conclusion}

In this work we have studied the effect of electron-hole asymmetry on the electronic structure of helical edge states in HgTe/HgCdTe quantum wells. We have obtained analytical expressions for the wave functions and the energy spectrum of helical states, the $g$-factor tensor, and the matrix elements of optical transitions between edge states with opposite spin in the framework of the electro-dipole mechanism. We have shown that in the presence of electron-hole asymmetry, the spectrum of edge states deviates from the linear one, and have found corrections of higher orders in the wave vector. It has been shown that electron-hole asymmetry has the greatest impact on the structure of helical states for a sharp boundary with vacuum, while in the case of a smooth boundary, for example, a contact between two insulators, its role is significantly reduced. Obtained results can be used in the analysis of magneto-transport phenomena and the edge photogalvanic effect in HgTe/HgCdTe quantum wells.

\acknowledgments

The author is grateful to S.A. Tarasenko for fruitful discussions. The author acknowledges financial support from the Russian Federation President Grant (project MK-2943.2019.2) and the ``Basis'' Foundation for the Advancement of Theoretical Physics and Mathematics.

\appendix
\section{Wave functions of edge states at $\D = 0$} \label{app}

In this section, we obtain analytic expressions for the wave functions $\psi_{k_y s}$ in the case $\D = 0$ and in the limit $\D/\B \ll 1$. The functions $\psi_{k_y s }$ are sought in the form~\eqref{wfs_ky_aniso}, where the set of equations for the functions $a_{1,2}$ and $b_{1,2}$ is given by Eq.~\eqref{sys}. We will look for a solution in the form $a_{1,2},~b_{1,2} \propto \e^{-\lambda x}$  with positive $\lambda$. Substituting this solution into the set of equations, we find the roots of the characteristic equation $\lambda_j$ and the corresponding vectors as functions of the energy $E$. It can be shown that the boundary condition $\psi_{k_y s}(x = 0) = 0$ can be satisfied only if $E = \A k_y/\sqrt{1 + k_0^2 l_2^2}$, see Eq.~\eqref{eps_ky_gamma} in the main text.

In the $l_1 \ll l_2$ limit we have:
\begin{equation}
\lambda_1 = \lambda_2 = \frac{1}{l_1}\:,~~~ \lambda_3 = \lambda_4^* = \frac{1}{l_2} - \mathrm{i} k_0 \sqrt{1 - \frac{E^2}{\delta_0^2}}\:.
\end{equation}
These equations show that for $|E| < |\delta_0|$  (the energy of the edge state lies in the bulk gap) only the oscillation period of the wave function varies, while the decay length does not change. For $|E| > |\delta_0|$, the decay length starts to increase, and finally, at $E = \pm \delta_0 \sqrt{1 + \delta_0^2/\gamma^2}$, it goes to infinity. As in the case $\gamma = 0$, this occurs at the point, where edge and bulk dispersion curves touch each other.

The final expressions for $a_{1,2}$ and $b_{1,2}$ read
\begin{eqnarray}
\label{a1a2b1b2}
a_1(x) &=& v_1 \left\{ \e^{-x/l_1} \cos \phi_1 - \e^{-x/l_2} \cos[k_0(E)x - \phi_1] \right\}\:, \nonumber \\
a_2(x) &=& v_1 \e^{-x/l_1} \cos \phi_1 - v_2 \e^{-x/l_2} \cos[k_0(E)x - \phi_1 - \phi_2] \:, \nonumber \\
b_1(x) &=& \e^{-x/l_1} \sin \phi_1 +v_1 v_2\e^{-x/l_2} \sin[k_0(E)x - \phi_1 - \phi_2]\:, \nonumber \\
b_2(x) &=& \e^{-x/l_1} \sin \phi_1  + \e^{-x/l_2} \sin[k_0(E)x - \phi_1]\:.
\end{eqnarray}
Here:
\begin{equation}
\label{k0_E}
k_0(E) = k_0 \sqrt{1 - \frac{E^2}{\delta_0^2}}\:,~~\tan 2 \phi_1 = \frac{k_0(E) l_2}{1 + k_y l_2}\:,
\end{equation}
\[
v_1 = \sqrt{\frac{|\delta_0|-E}{|\delta_0|+E}}\:,~ v_2 = \sqrt{\frac{|\delta_0| \sqrt{1 + k_0^2 l_2^2} - E}{|\delta_0|\sqrt{1 + k_0^2 l_2^2} + E}}\:,
\]
\[
\tan \phi_2 = \frac{E k_0 l_2}{\sqrt{\delta_0^2-E^2} \sqrt{1+k_0^2l_2^2}}\:.
\]

As shown in Sec.~\ref{low_D}, in the limit $\D/\B \ll 1$ Eqs.~\eqref{sys} has the same form as at $\D = 0$, but with $l_2$ that depends on energy according to Eq.~\eqref{l2_E}. Thus, to find the wave functions of the edges states in the limit $\D/\B \ll 1$, one should substitute the $l_2$ length in Eqs.~\eqref{a1a2b1b2} and \eqref{k0_E} with its expression Eq.~\eqref{l2_E}.


\end{document}